\documentclass[conference]{IEEEtran}
\IEEEoverridecommandlockouts

\usepackage{url}
\usepackage{cite}
\usepackage{amsmath,amssymb,amsfonts}
\usepackage{dblfloatfix}
\usepackage{placeins}
\usepackage{algorithmic}
\usepackage{graphicx}
\usepackage{textcomp}
\usepackage{xcolor}
\usepackage{colortbl}
\usepackage[colorinlistoftodos]{todonotes}
\usepackage{multirow}
\usepackage{multicol}
\usepackage{enumitem}
\usepackage{hhline}
\usepackage{threeparttable}
\usepackage{booktabs}
\def\BibTeX{{\rm B\kern-.05em{\sc i\kern-.025em b}\kern-.08em
    T\kern-.1667em\lower.7ex\hbox{E}\kern-.125emX}}
\begin{document}
\title{Voice Anonymization Made Simple: Training-Free Anonymization with Projected Classifier-Free Guidance
}

\author{
\IEEEauthorblockN{
  Xiang Shi\IEEEauthorrefmark{1},
  Han Zhu\IEEEauthorrefmark{2},
  Ming Li\IEEEauthorrefmark{3},
  Xiaoxiao Miao\IEEEauthorrefmark{1}
}
\IEEEauthorblockA{
  \IEEEauthorrefmark{1}Duke Kunshan University\\
  \IEEEauthorrefmark{2}Xiaomi Corporation\\
  \IEEEauthorrefmark{3}School of Artificial Intelligence,
  The Chinese University of Hong Kong, Shenzhen
}
\thanks{xiang.shi@dukekunshan.edu.cn, xiaoxiao.miao@dukekunshan.edu.cn.

Corresponding author: Xiaoxiao Miao.}
}
\IEEEaftertitletext{\vspace{-2\baselineskip}}

\maketitle

\begin{abstract}
Voice anonymization aims to conceal speaker identity while preserving
linguistic and paralinguistic information, yet many existing methods require
dedicated training. We propose a training-free, two-stage anonymization method
for pretrained flow-matching voice conversion systems. In
the condition-construction stage, the source content representation is
regenerated under a randomly selected speaker context to reduce residual
source-speaker information while preserving linguistic content. In the
speech-generation stage, the source speech is used to identify the generation
direction associated with the original speaker, and the source-aligned
component is suppressed while useful content guidance is retained. The proposed
method modifies only inference-time conditioning and guidance, without
updating any pretrained parameters. Experiments on VoicePrivacy 2026 Track~1
show that the two-stage method improves speaker privacy while maintaining
competitive word error rate and emotion recognition performance\footnote{Full code and audio samples can be found at
\url{https://github.com/xs125-stack/voice_anon_simple}}.
\end{abstract}


\section{Introduction}

Speech conveys not only linguistic content but also personal information, including identity, health status, emotional state, geographical background, and social characteristics \cite{backstrom2025privacy}. As speech is increasingly used for authentication, analytics, human--computer interaction, and generative applications, information beyond the intended purpose of collection may be inferred from recorded utterances. Voice privacy therefore concerns limiting unauthorized access to, or inference of, personal information conveyed through speech \cite{backstrom2025privacy}.

This work focuses on voice anonymization, which aims to prevent reliable re-identification of the source speaker while preserving information required by downstream applications, such as linguistic content and prosodic attributes \cite{shamsabadi2022differentially}. The task was formalized by the VoicePrivacy Challenge, which provides common datasets, baselines, and evaluation protocols \cite{tomashenko2020introducing,tomashenko2022voiceprivacy}. Subsequent editions introduced increasingly stronger attacker models, including semi-informed attackers \cite{tomashenko2022voiceprivacy2022challengeevaluation,panariello2024voiceprivacy}, emotion-preservation evaluation \cite{tomashenko2024voiceprivacy,tomashenko2026third}, and, most recently, stronger ASV attackers pretrained on voice-converted speech together with cross-gender and multilingual evaluation \cite{arefeen2026dast,miao2026voiceprivacy}.

Voice anonymization methods broadly follow signal-processing or neural
synthesis approaches. Signal-processing methods modify acoustic properties such
as formants or speaking rate, but may distort speech and remain vulnerable to
strong attackers \cite{gupta2020design,dubagunta2020adjustable,
mcadams1984spectral,vpc22tsm,tavi2022improving,
tomashenko2022voiceprivacy,panariello2024voiceprivacy}.

Neural synthesis methods reconstruct speech from transformed or disentangled
representations. ASR+TTS removes speaker identity through text resynthesis but
also discards paralinguistic information \cite{xinyuan24_spsc}, while more
advanced systems preserve prosodic attributes or disentangle content, prosody,
and speaker information before anonymization
\cite{meyer2023prosody,fang2019speaker,shamsabadi2022differentially,
mawalim2022speaker,champion2023anonymizing,yao2024musa,miao22_odyssey,
miao2023language,yao2024distinctive}. Other approaches anonymize general speech
representations or discrete speech tokens without explicit speaker embeddings
\cite{cai2024privacy,das2024comparing,tang2025sef,panariello2024speaker}.

Diffusion and flow-matching models have recently emerged as powerful generative frameworks for voice anonymization. DiffVC+ adapts diffusion-based voice conversion for anonymization \cite{huang24_interspeech}; F3-VA generates anonymized speaker embeddings using flow matching before conditional speech reconstruction \cite{xiao2026anonymization}; and DiffAnon reconstructs codec representations from semantic, prosodic, and speaker conditions with classifier-free guidance to control prosody preservation \cite{ulgen2026diffanon}. These methods demonstrate that diffusion and flow matching enable high-quality and controllable speech anonymization.

However, most existing approaches require anonymization-specific training,
additional learned components, or retraining when applied to a different voice
conversion model. This raises a broader question: can a pretrained flow-matching-based voice conversion model be converted into a
voice anonymizer only by changing its inference procedure?

We address this question with a training-free, two-stage anonymization
framework for pretrained generative voice conversion systems. The main idea is
to intervene at both the representation and generation levels. During
condition construction, the source content representation is regenerated under
an external speaker context to reduce its dependence on the original speaker.
During speech generation, the source speech is used to identify the generation
direction associated with the source identity, and the source-aligned component
is suppressed while useful content guidance is retained. Both operations are
performed only at inference time and require no parameter updates.

We implement this framework on Seed-VC v2, a state-of-the-art zero-shot voice
conversion system \cite{liu2024seedvc}. Seed-VC v2 follows a two-stage
generation process: it first constructs the conditions required for conversion
and then synthesizes speech through conditional flow matching. This structure
makes it suitable for applying our two interventions separately. In the first
stage, a randomly selected target utterance guides the reconstruction of the
source-content condition. In the second stage, source-aware projected
classifier-free guidance steers generation away from the original speaker.

\begin{figure*}[t]
\centering
\includegraphics[width=0.94\textwidth]{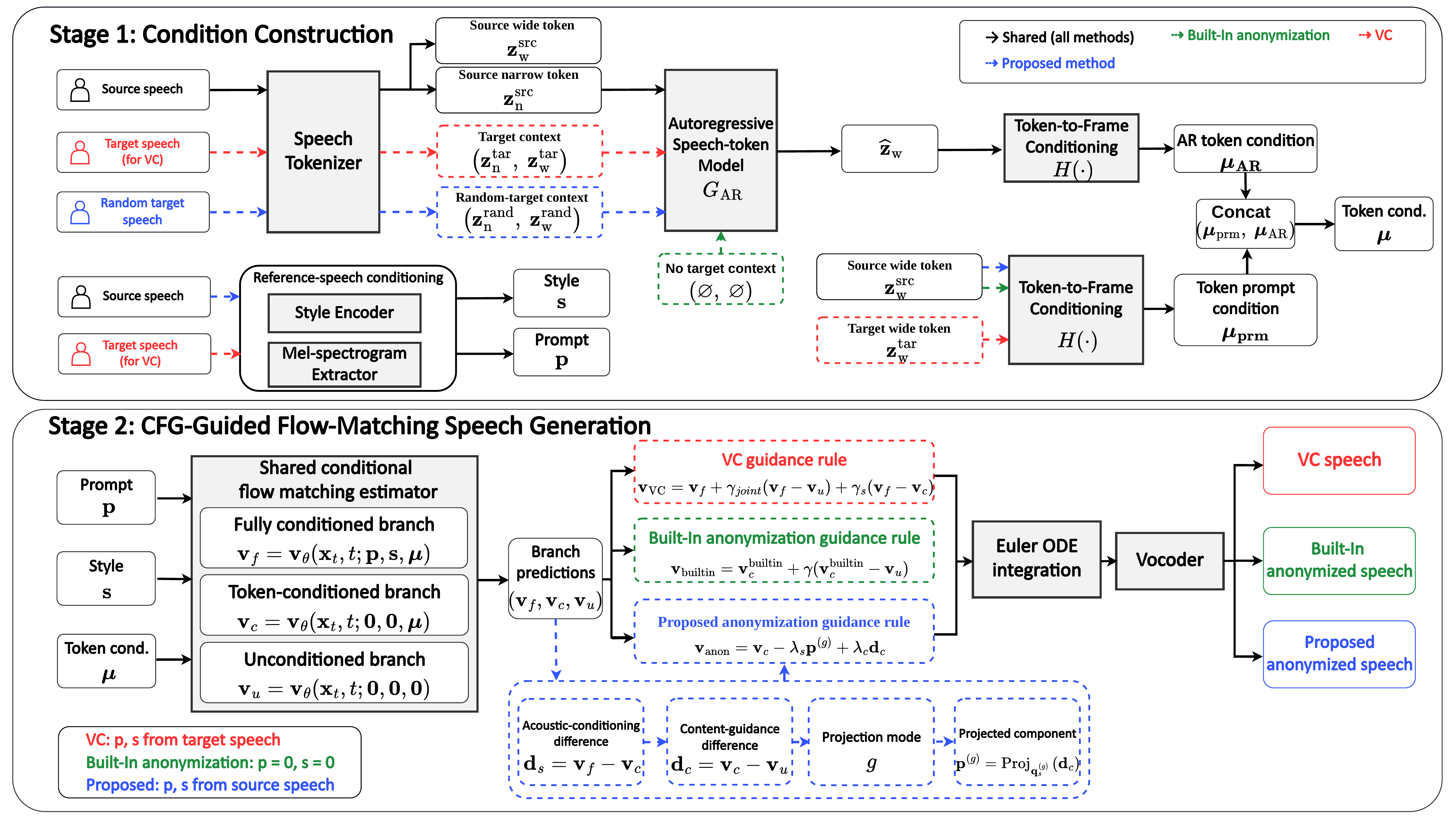}
\caption{Overview of standard voice conversion, built-in anonymization, and
the proposed training-free method within the shared Seed-VC v2 backbone. The
proposed method modifies condition construction to reduce source-speaker
dependence and modifies generation guidance to suppress source-aligned
information. Detailed condition and guidance definitions are introduced in
Sections~\ref{sec:pre} and \ref{sec:method}.}
\label{fig:proposed_method}
\end{figure*}

Experiments under the VoicePrivacy 2026 Track 1 (English) protocol show that the proposed training-free two-stage intervention achieves a favorable privacy–utility trade-off, improving speaker privacy while maintaining competitive linguistic and emotion utility compared with the baselines.

\section{Preliminaries}
\label{sec:pre}

\subsection{Voice Conversion in Seed-VC v2}

Figure~\ref{fig:proposed_method} shows the released Seed-VC v2 pipeline.
Seed-VC is a zero-shot voice conversion system that transfers a source
utterance to an unseen reference speaker while preserving linguistic content
\cite{liu2024seedvc}. The published Seed-VC model uses continuous semantic
features, whereas v2 inserts an autoregressive (AR) speech-token model before
the conditional flow-matching acoustic model.\footnote{\url{https://github.com/Plachtaa/seed-vc/blob/main/inference_v2.py}}
Both versions share the same basic idea: the reference speaker is represented
by both a global timbre vector and a complete reference segment that provides
in-context information \cite{liu2024seedvc}. In
Figure~\ref{fig:proposed_method}, black arrows denote shared computation and
red arrows denote standard VC.

\noindent\textbf{Stage~1: Condition construction.}
This stage constructs the complete token condition
\(\boldsymbol{\mu}\), the acoustic prompt \(\mathbf{p}\), and the global style
embedding \(\mathbf{s}\). The complete token condition is not simply a source
content representation. It contains a prompt-side part aligned with the
reference speech and an AR-generated part for the content to be synthesized.

The source and target utterances are processed by pretrained narrow- and
wide-band tokenizers.\footnote{\url{https://github.com/Plachtaa/ASTRAL-quantization}}
The source narrow-band tokens provide the linguistic sequence, while the target
narrow- and wide-band tokens serve as AR context and prefix. The AR model
therefore generates a target-conditioned wide-band sequence for the source
portion,
\begin{equation}
\widehat{\mathbf{z}}_{\mathrm{w}}
=
G_{\mathrm{AR}}
\left(
\mathbf{z}^{\mathrm{src}}_{\mathrm{n}};
\mathbf{z}^{\mathrm{tar}}_{\mathrm{n}},
\mathbf{z}^{\mathrm{tar}}_{\mathrm{w}}
\right),
\label{eq:seedvc_ar}
\end{equation}
which is converted to the Mel-frame rate as
\(\boldsymbol{\mu}_{\mathrm{AR}}=H(\widehat{\mathbf{z}}_{\mathrm{w}})\).
Here, \(\boldsymbol{\mu}_{\mathrm{AR}}\) describes the source content to be
generated, but in a token form already influenced by the target context.

The target wide-band tokens are regulated separately to form the token-prompt
condition
\(\boldsymbol{\mu}_{\mathrm{prm}}=H(\mathbf{z}^{\mathrm{tar}}_{\mathrm{w}})\).
The two parts are then concatenated along time as
\(\boldsymbol{\mu}=\operatorname{Concat}_{t}
(\boldsymbol{\mu}_{\mathrm{prm}},\boldsymbol{\mu}_{\mathrm{AR}})\).
They are regulated separately because the prompt region and generated region
must be aligned to different Mel lengths before concatenation.

The same target utterance also provides the acoustic prompt and global style
embedding, \(\mathbf{p}=M(x^{\mathrm{tar}})\) and
\(\mathbf{s}=E_{\mathrm{s}}(x^{\mathrm{tar}})\), where \(M(\cdot)\) is
Mel-spectrogram extraction and \(E_{\mathrm{s}}(\cdot)\) is a pretrained
CAM++ encoder \cite{wang23ha_interspeech}. Their roles are complementary:
\(\boldsymbol{\mu}_{\mathrm{prm}}\) tells the model what is spoken in the
reference segment, \(\mathbf{p}\) shows how that segment is realized
acoustically, \(\mathbf{s}\) gives the overall speaker and timbre, and
\(\boldsymbol{\mu}_{\mathrm{AR}}\) specifies the content to generate next.

Together, these conditions form an aligned in-context reference and consistently direct standard VC toward the target speaker \cite{liu2024seedvc}.

\noindent\textbf{Stage~2: Speech generation.}
The conditional flow-matching model uses
\((\boldsymbol{\mu},\mathbf{p},\mathbf{s})\) to generate the converted
Mel-spectrogram. Its velocity network is a diffusion transformer (DiT)
\cite{peebles2023dit}. At flow time \(t\) and state \(\mathbf{x}_{t}\), the
estimator
\(\mathbf{v}_{\theta}(\mathbf{x}_{t},t;\mathbf{p},\mathbf{s},
\boldsymbol{\mu})\) defines the ODE
\(d\mathbf{x}_{t}/dt=\mathbf{v}_{\theta}\), which is integrated from Gaussian
noise by a fixed-step Euler solver.

For multi-condition classifier-free guidance (CFG), the estimator is evaluated
under three condition sets,
\begin{equation}
\begin{aligned}
\mathbf{v}_{f}
&=
\mathbf{v}_{\theta}
\left(
\mathbf{x}_{t},t;
\mathbf{p},
\mathbf{s},
\boldsymbol{\mu}
\right),\\
\mathbf{v}_{c}
&=
\mathbf{v}_{\theta}
\left(
\mathbf{x}_{t},t;
\mathbf{0},
\mathbf{0},
\boldsymbol{\mu}
\right),\\
\mathbf{v}_{u}
&=
\mathbf{v}_{\theta}
\left(
\mathbf{x}_{t},t;
\mathbf{0},
\mathbf{0},
\mathbf{0}
\right),
\end{aligned}
\label{eq:seedvc_branches}
\end{equation}
giving the fully conditioned, token-conditioned, and unconditional
predictions. Standard VC combines them as
\begin{equation}
\mathbf{v}_{\mathrm{VC}}
=
\mathbf{v}_{f}
+
\gamma_{\mathrm{joint}}
\left(
\mathbf{v}_{f}-\mathbf{v}_{u}
\right)
+
\gamma_{\mathrm{s}}
\left(
\mathbf{v}_{f}-\mathbf{v}_{c}
\right),
\label{eq:seedvc_vc_rule}
\end{equation}
where \(\gamma_{\mathrm{joint}}\) scales the joint guidance contributed by
all conditions, including both token/content and reference-speaker
information. In contrast, \(\gamma_{\mathrm{s}}\) scales the additional
guidance introduced by the acoustic prompt and style embedding. 
The Euler solver produces a converted Mel-spectrogram, which a pretrained
BigVGAN v2 vocoder \cite{lee2023bigvgan} converts into waveform audio.

\subsection{Built-in Anonymization in Seed-VC v2}

Seed-VC v2 also provides a training-free built-in anonymization mode, shown by
the green path in Figure~\ref{fig:proposed_method}. It keeps the same two-stage
backbone but removes explicit target-timbre conditioning.

\noindent\textbf{Stage~1: Condition construction.}
The AR model receives the source narrow-band tokens but no target context or
prefix,
\begin{equation}
\widehat{\mathbf{z}}_{\mathrm{w}}^{\mathrm{builtin}}
=
G_{\mathrm{AR}}
\left(
\mathbf{z}^{\mathrm{src}}_{\mathrm{n}};
\varnothing,
\varnothing
\right),
\qquad
\boldsymbol{\mu}_{\mathrm{AR}}^{\mathrm{builtin}}
=
H\left(
\widehat{\mathbf{z}}_{\mathrm{w}}^{\mathrm{builtin}}
\right),
\label{eq:builtin_mu_ar}
\end{equation}
where \(\varnothing\) denotes an absent AR context or prefix.

The released wrapper still constructs a token-prompt prefix from
\texttt{target\_audio\_path}. However, because the acoustic prompt and style
embedding are set to zero, this prefix no longer forms an aligned in-context
reference pair. Our preliminary tests found little difference between using a
source-derived or target-derived prefix. We therefore use the source utterance
by default, giving
\(\boldsymbol{\mu}_{\mathrm{prm}}^{\mathrm{builtin}}
=H(\mathbf{z}^{\mathrm{src}}_{\mathrm{w}})\).
The complete built-in token condition is
\(\boldsymbol{\mu}_{\mathrm{builtin}}
=\operatorname{Concat}_{t}
(\boldsymbol{\mu}_{\mathrm{prm}}^{\mathrm{builtin}},
\boldsymbol{\mu}_{\mathrm{AR}}^{\mathrm{builtin}})\),
with \(\mathbf{p}=\mathbf{0}\) and \(\mathbf{s}=\mathbf{0}\).

\noindent\textbf{Stage~2: Speech generation.}
Generation therefore uses only the complete token condition. The estimator is
evaluated under the token-conditioned and unconditional settings,
\begin{equation}
\mathbf{v}_{c}^{\mathrm{builtin}}
=
\mathbf{v}_{\theta}
\left(
\mathbf{x}_{t},t;
\mathbf{0},
\mathbf{0},
\boldsymbol{\mu}_{\mathrm{builtin}}
\right),
\qquad
\mathbf{v}_{u}
=
\mathbf{v}_{\theta}
\left(
\mathbf{x}_{t},t;
\mathbf{0},
\mathbf{0},
\mathbf{0}
\right),
\label{eq:builtin_branches}
\end{equation}
and combined as
\begin{equation}
\mathbf{v}_{\mathrm{builtin}}
=
\mathbf{v}_{c}^{\mathrm{builtin}}
+
\gamma_{}
\left(
\mathbf{v}_{c}^{\mathrm{builtin}}
-
\mathbf{v}_{u}
\right),
\label{eq:builtin_vector_field}
\end{equation}
where \(\gamma_{}\) is the intelligibility CFG scale. The Euler solver
integrates this field, and the vocoder renders the anonymized waveform.

\section{Proposed Method}
\label{sec:method}

The built-in Seed-VC anonymization mode already provides a training-free way to
anonymize speech without updating the pretrained model. However, it still leaves
room for improvement in both stages. In Stage~1, the target-related AR inputs are
simply removed, so the model does not exploit its ability to regenerate the
source content under an alternative speaker context. As a result, the token
condition is still largely determined by the source input and may carry residual
source-speaker information entangled with the content. In Stage~2, the guidance
rule \(\mathbf{v}_{c}^{\mathrm{builtin}}+\gamma_{c}(\mathbf{v}_{c}^{\mathrm{builtin}}-\mathbf{v}_{u})\)
reduces the direct use of source acoustic conditions, but it neither identifies
and removes the source-related information carried in the token guidance nor
steers the generation trajectory away from the source speaker. These
observations motivate our two modifications: random-target conditioning in
Stage~1, which regenerates the source content under an independently sampled
speaker context to weaken residual source-speaker information, and
source-conditioning-aware projected CFG in Stage~2, which actively suppresses
source-aligned directions during generation.
The two modifications are highlighted as the blue path in Figure~\ref{fig:proposed_method}.

\subsection{Stage 1: Random-Target-Conditioned Token and Source-Acoustic Condition Construction}
\label{sec:condition_construction}

In Stage~1, we change only the AR-generated part of the token condition. As
shown in the upper panel of Figure~\ref{fig:proposed_method}, the complete token
condition contains two parts: the token-prompt condition
\(\boldsymbol{\mu}_{\mathrm{prm}}\) and the AR token condition
\(\boldsymbol{\mu}_{\mathrm{AR}}\). The former describes the prompt region,
whereas the latter describes the source content to be generated.

To reduce source-speaker information in the generated source-content tokens, we
sample an independent random-target utterance and use its narrow- and wide-band
tokens as the AR context and prefix. The source narrow-band tokens still provide
the linguistic sequence to be converted. The shared AR model therefore predicts
the source-portion wide-band tokens as
\begin{equation}
\widehat{\mathbf{z}}_{\mathrm{w}}
=
G_{\mathrm{AR}}
\left(
\mathbf{z}^{\mathrm{src}}_{\mathrm{n}};
\mathbf{z}^{\mathrm{rand}}_{\mathrm{n}},
\mathbf{z}^{\mathrm{rand}}_{\mathrm{w}}
\right).
\label{eq:ar_conversion}
\end{equation}
The generated sequence is then converted to the Mel-frame rate as
\(\boldsymbol{\mu}_{\mathrm{AR}}=H(\widehat{\mathbf{z}}_{\mathrm{w}})\).
Thus, \(\boldsymbol{\mu}_{\mathrm{AR}}\) still carries the source linguistic
content, but its token realization is guided by an unrelated speaker context.

The token-prompt condition is constructed separately from the source
wide-band tokens as
\(\boldsymbol{\mu}_{\mathrm{prm}}=H(\mathbf{z}^{\mathrm{src}}_{\mathrm{w}})\).
We deliberately keep \(\boldsymbol{\mu}_{\mathrm{prm}}\) source-derived because
the same source utterance also provides the acoustic prompt
\(\mathbf{p}\). This keeps the token description of the prompt region aligned
with its Mel-spectrogram. If the token-prompt condition were instead taken from
the random target while the acoustic prompt remained source-derived, the model
would receive random-target prompt tokens together with a source acoustic
example. This prompt mismatch differs from the aligned conditioning used during
Seed-VC training. In our preliminary tests, such a mismatch sometimes improved
speaker separation, but it also made generation unstable and substantially degraded audio quality and increased WER. We therefore use the source utterance for
both the token-prompt condition and the acoustic prompt.

The complete token condition is formed by temporal concatenation as
\(\boldsymbol{\mu}=\operatorname{Concat}_{t}
(\boldsymbol{\mu}_{\mathrm{prm}},\boldsymbol{\mu}_{\mathrm{AR}})\).
This operation is shown explicitly by the \emph{Concat} block in
Figure~\ref{fig:proposed_method}. The two parts are first regulated to their own
Mel lengths and are concatenated only afterward.

In parallel, the source utterance is passed through the same Mel-spectrogram
extractor and CAM++ style encoder used by Seed-VC, giving
\(\mathbf{p}=M(x^{\mathrm{src}})\) and
\(\mathbf{s}=E_{\mathrm{s}}(x^{\mathrm{src}})\), where \(M(\cdot)\) denotes
prompt-Mel extraction and \(E_{\mathrm{s}}(\cdot)\) denotes the style
encoder. The source-derived
\(\boldsymbol{\mu}_{\mathrm{prm}}\) and \(\mathbf{p}\) form an aligned prompt
example, while \(\mathbf{s}\) describes the overall source timbre. However,
these source conditions are not used as desired output attributes. They are
retained only to expose the effect of source-acoustic conditioning in Stage~2.

Because the complete token condition retains a source prompt part and AR
generation starts from source tokens, Stage~1 reduces but cannot guarantee the removal of all source-speaker information.

\subsection{Stage 2: Source-Conditioning-Aware Projected Classifier-Free Guidance for Conditional Flow-Matching Speech Generation}
\label{sec:projected_cfg}

In Stage~2, we use the source prompt and style embedding only to identify the
generation direction associated with the source speaker. At each ODE step, the
difference between the fully conditioned and token-conditioned predictions
shows how the source acoustic conditions change the current generation state.
The difference between the token-conditioned and unconditional predictions
shows the effect of the complete token condition
\(\boldsymbol{\mu}\), which contains both
\(\boldsymbol{\mu}_{\mathrm{prm}}\) and
\(\boldsymbol{\mu}_{\mathrm{AR}}\). We project this token-guidance direction
onto the source-conditioning direction and suppress the aligned component.

Accordingly, the source prompt \(\mathbf{p}\) and style embedding
\(\mathbf{s}\) serve only as a reference. They identify what should be
suppressed rather than what should be reconstructed. The remaining token
guidance continues to support linguistic generation, so sampling starts from
the token-conditioned field \(\mathbf{v}_{c}\), not from the fully conditioned
field \(\mathbf{v}_{f}\).

Concretely, at every ODE step, the shared estimator is evaluated under the same
three condition sets as in Eq.~\eqref{eq:seedvc_branches}, using the
\((\boldsymbol{\mu},\mathbf{p},\mathbf{s})\) constructed in Stage~1 to obtain
\(\mathbf{v}_{f}\), \(\mathbf{v}_{c}\), and \(\mathbf{v}_{u}\). We then form
\begin{equation}
\mathbf{d}_{s}
=
\mathbf{v}_{f}-\mathbf{v}_{c},
\qquad
\mathbf{d}_{c}
=
\mathbf{v}_{c}-\mathbf{v}_{u},
\label{eq:guidance_directions}
\end{equation}
where \(\mathbf{d}_{s}\) is the source-conditioning direction contributed by
the source acoustic prompt and style embedding, and \(\mathbf{d}_{c}\) is the
guidance direction contributed by the complete token condition. Their
alignment identifies the part of token guidance that changes the acoustic state
in a way consistent with reintroducing source conditioning. We do not assume
that \(\mathbf{d}_{s}\) is a pure speaker direction or that
\(\mathbf{d}_{c}\) is a pure content direction.

Both \(\mathbf{d}_{s}\) and \(\mathbf{d}_{c}\) have shape \(C\times T\), where
\(C\) is the feature dimension and \(T\) is the number of frames. The projection
reference \(\mathbf{q}^{(g)}_{s,t}\) is built from
\(\mathbf{d}_{s}\) at one of three temporal granularities,
\begin{equation}
\mathbf{q}^{(g)}_{s,t}
=
\begin{cases}
\mathbf{d}_{s,t},
& g=\mathrm{token},\\[4pt]
\dfrac{1}{T}\displaystyle\sum_{\tau=1}^{T}\mathbf{d}_{s,\tau},
& g=\mathrm{global},\\[10pt]
\dfrac{1}{K}\displaystyle\sum_{\tau\in\mathcal{W}_{t}}\mathbf{d}_{s,\tau},
& g=\mathrm{smooth},
\end{cases}
\label{eq:source_reference}
\end{equation}
where \emph{Token} uses the source-conditioning vector at the same frame,
\emph{Global} uses one utterance-level mean, and \emph{Smooth} uses a local
sliding-window mean over
\(\mathcal{W}_{t}=\{t-10,\dots,t+10\}\).\footnote{\(K=21\): the current frame
plus the ten preceding and ten following frames, with reflection padding at
utterance boundaries. We therefore refer to this smooth variant as
\emph{Smooth-21} in the experiments.} These choices respectively preserve frame-level
variation, remove local variation, or provide an intermediate smoothed
reference.

The component of token guidance aligned with the selected reference is
\begin{equation}
\mathbf{p}^{(g)}_{t}
=
\frac{
\left\langle
\mathbf{d}_{c,t},
\mathbf{q}^{(g)}_{s,t}
\right\rangle
}{
\max\!\left(
\left\|\mathbf{q}^{(g)}_{s,t}\right\|_{2}^{2},
\epsilon
\right)
}
\mathbf{q}^{(g)}_{s,t}.
\label{eq:projection}
\end{equation}
The vector being projected is always \(\mathbf{d}_{c,t}\); only the temporal
construction of the reference changes across modes. We treat
\(\mathbf{p}^{(g)}\) as a proxy for source-speaker information remaining in the
token pathway.

The proposed guidance rule removes this component and then reinforces the token
guidance,
\begin{equation}
\mathbf{v}_{\mathrm{anon}}
=
\mathbf{v}_{c}
-
\lambda_{s}\,\mathbf{p}^{(g)}
+
\lambda_{c}\,\mathbf{d}_{c},
\label{eq:final_vector_field}
\end{equation}
where \(\lambda_{s}\) controls source-aligned suppression and
\(\lambda_{c}\) controls token guidance. Using \(\mathbf{v}_{c}\) as the base
field prevents the source prompt and style from directly reconstructing the
source voice, while the difference between \(\mathbf{v}_{f}\) and
\(\mathbf{v}_{c}\) still indicates which direction should be removed. The shared Euler solver then integrates
\(\mathbf{x}_{t_{k+1}}=\mathbf{x}_{t_k}
+\Delta t_k\,\mathbf{v}_{\mathrm{anon}}(\mathbf{x}_{t_k},t_k)\),
and the pretrained vocoder renders the final waveform.

\begin{figure*}[t]
    \centering
    \vspace{-2mm}
    \includegraphics[
        width=0.88\textwidth,
        trim=0 18 10 10,
        clip
    ]{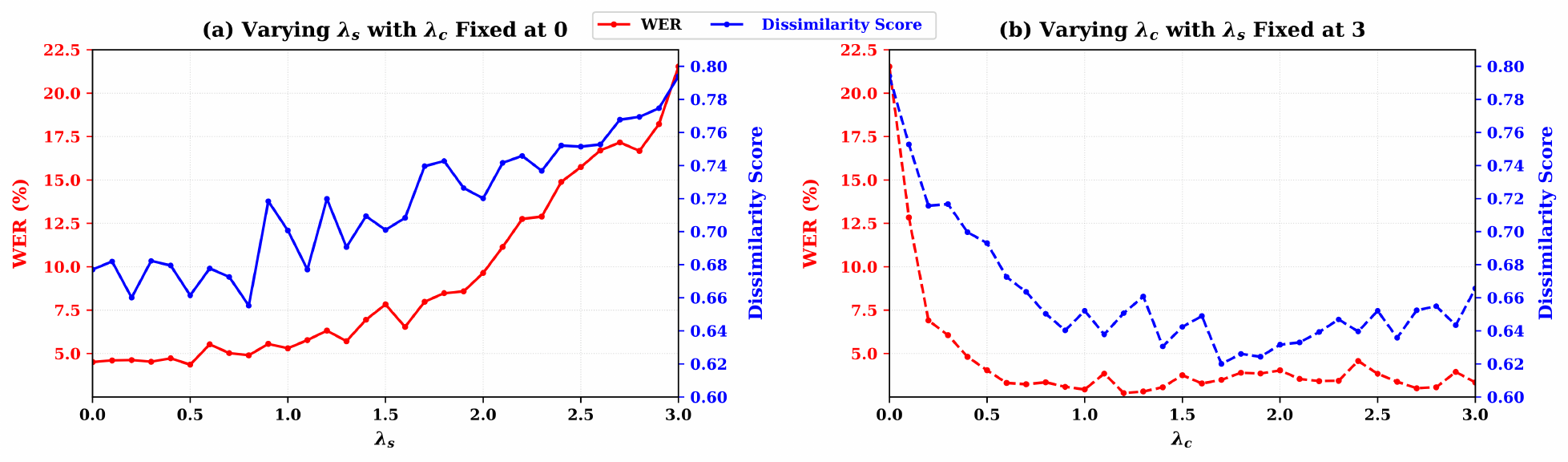}
    \vspace{-2mm}
    \caption{Effects of \(\lambda_s\) and \(\lambda_c\) on the
    30-utterance screening subset. A higher dissimilarity score indicates
    lower source-speaker similarity, and WER is computed at the corpus level.}
    \label{fig:scale_sweeps}
    \vspace{-3mm}
\end{figure*}

\section{Experiments}

Our experiments proceed in three steps. First, we construct a small screening
subset containing 30 utterances selected from the VPC LibriSpeech development clean set \cite{librispeech}, denoted as \emph{Libri\_dev}. \footnote{The screening subset contains 30 speakers, with one long and clearly spoken utterance selected per speaker. It is used only for quick screening; empirically, the selected settings generalize well to the full development and evaluation sets, although tuning on the full development set could potentially yield better settings and would make the dense \((\lambda_s,\lambda_c)\) sweep substantially more expensive.} On this subset, we test whether the two coefficients
in Eq.~\eqref{eq:final_vector_field} behave as intended: increasing
\(\lambda_s\) should strengthen anonymization by removing more
source-aligned guidance, whereas increasing \(\lambda_c\) should strengthen
content guidance and reduce WER. Second, after confirming these roles, we use
the same subset to search for the best \((\lambda_s,\lambda_c)\) pair for each
combination of random-target condition construction and projection mode.
Third, with the selected scale pairs fixed, we evaluate all systems on the full
VPC 2026 Track~1 development and evaluation sets. These full-set experiments
serve as ablation studies for the random-target-conditioned construction of
\(\boldsymbol{\mu}_{\mathrm{AR}}\) introduced in
Section~\ref{sec:condition_construction} and the projected guidance proposed in
Section~\ref{sec:projected_cfg}.

\subsection{Evaluation Protocol, Datasets, and Settings}

We follow the VoicePrivacy Challenge 2026 Track~1 (English) evaluation protocol
\cite{miao2026voiceprivacy} and use the challenge-provided datasets, attack
models, and automatic speech recognition (ASR) and speech emotion recognition
(SER) evaluation models without modification. Speaker privacy is measured by
equal error rate (EER) under lazy-informed and semi-informed attacks, where a
higher value indicates stronger protection against speaker re-identification.
Linguistic-content preservation is measured by word error rate (WER, lower is
better), and emotion preservation is measured by unweighted average recall
(UAR, higher is better).

The random-target pool introduced in Section~\ref{sec:method} contains one
short utterance (3--6\,s) from each of 500 speakers in the LibriTTS
\texttt{train-other-500} subset \cite{zen2019libritts}. For every source
utterance used by a random-target-conditioned system, one target utterance is
sampled from this pool with replacement. The sampled target utterance provides the narrow-band context and wide-band
prefix used to construct the AR-generated part
\(\boldsymbol{\mu}_{\mathrm{AR}}\). The token-prompt part
\(\boldsymbol{\mu}_{\mathrm{prm}}\) remains source-derived so that it stays
aligned with the source acoustic prompt. A fixed random seed is used, and the
same source--target assignment is retained across all projection modes and
scale settings to ensure a controlled comparison. \footnote{We also performed preliminary checks with several alternative random seeds and source--target assignments and observed only small changes. These checks suggest that using an unrelated target context is more important than the exact target identity for weakening source-speaker information in
\(\boldsymbol{\mu}_{\mathrm{AR}}\).}

\subsection{Results}
\noindent\textbf{Verifying the Roles of the Two Guidance Scales.}
We take the subset \emph{Libri\_dev} to verify whether the two coefficients in
Eq.~\eqref{eq:final_vector_field} have the intended effects. All configurations
use the same source utterances. For random-target-conditioned configurations, the same fixed
source--target pairs are used throughout the sweep. Only the AR-generated part
\(\boldsymbol{\mu}_{\mathrm{AR}}\) changes with the sampled target;
\(\boldsymbol{\mu}_{\mathrm{prm}}\) remains source-derived in all
configurations.
Privacy is measured on this subset using the dissimilarity score
\[
A
=
1-
\frac{1}{N}
\sum_{i=1}^{N}
\cos
\left(
\mathbf{e}^{\mathrm{src}}_{i},
\mathbf{e}^{\mathrm{anon}}_{i}
\right),
\qquad N=30,
\]
where the embeddings are extracted using the same speaker model as the VPC ASV
pipeline. A higher value indicates that the anonymized speech is farther from
the source speaker. Utility is measured by corpus-level WER over the same
30 utterances using the VPC ASR model.

Figure~\ref{fig:scale_sweeps} varies one coefficient while keeping the other
fixed. \footnote{In Fig.~\ref{fig:scale_sweeps}, \(\lambda_s=3\) is used as the upper sweep bound for diagnostic analysis, not as the selected operating point.} As expected, increasing \(\lambda_s\) raises the dissimilarity score,
although excessive suppression also increases WER. Increasing
\(\lambda_c\), in contrast, lowers WER by strengthening token guidance, but it
also reduces the dissimilarity score. These opposite trends confirm the roles
assigned to the two terms: \(\lambda_s\) mainly controls source-speaker
suppression, whereas \(\lambda_c\) mainly controls linguistic-content
preservation.

\begin{table}[!t]
\caption{Development-set ablation results. RT means that
\(\boldsymbol{\mu}_{\mathrm{AR}}\) uses random-target context.
Higher EER and UAR are better, while lower WER is better.
The highest EER is shaded.}
\label{tab:track1-ablation}
\centering

\resizebox{0.8\columnwidth}{!}{%
\begin{tabular}{@{}lcccccc@{}}
\toprule
\textbf{System}
&
\textbf{RT}
&
\(\boldsymbol{\lambda_s}\)
&
\(\boldsymbol{\lambda_c}\)
&
\textbf{EER}
&
\textbf{WER}
&
\textbf{UAR}
\\
\midrule

Built-in
&
&
0.7
&
0.7
&
33.15
&
4.18
&
41.85
\\

\midrule

Global
&
&
0.9
&
2.8
&
33.90
&
4.20
&
39.21
\\

Smooth-21
&
&
1.0
&
3.0
&
34.20
&
4.39
&
40.35
\\

Token
&
&
0.2
&
2.4
&
33.48
&
4.22
&
40.30
\\

\midrule
Global
&
$\checkmark$
&
0.9
&
2.8
&
34.44
&
4.72
&
38.62
\\

Smooth-21
&
$\checkmark$
&
1.0
&
3.0
&
34.79
&
4.47
&
39.06
\\

Token
&
$\checkmark$
&
0.2
&
2.4
&
33.80
&
4.60
&
39.43
\\

\midrule

Global
&
$\checkmark$
&
1.2
&
2.9
&
34.61
&
4.42
&
40.13
\\

Smooth-21
&
$\checkmark$
&
1.1
&
2.8
&
\cellcolor{gray!15}{35.09}
&
4.52
&
40.38
\\

Token
&
$\checkmark$
&
0.1
&
1.3
&
34.72
&
4.36
&
39.95
\\

\bottomrule
\end{tabular}%
}
\end{table}

\begin{table}[!t]
\caption{Track~1 comparison with the official baselines. Privacy is measured
by EER (\%) under lazy-informed (L) and semi-informed (S) attacks, and utility
by WER (\%) and UAR (\%). Built-in anon.\ uses
\((\lambda_s,\lambda_c)=(0.7,0.7)\) and Smooth-21 uses \((1.1,2.8)\);
\(\dagger\) marks systems using random-target context for
\(\boldsymbol{\mu}_{\mathrm{AR}}\). The maximum value in each EER column is
shaded.}
\label{tab:track1-semi}
\centering
\scriptsize
\setlength{\tabcolsep}{3pt}
\renewcommand{\arraystretch}{1.1}
\begin{tabular}{@{}l|cccc|cccc@{}}
\toprule
\multirow{2}{*}{\textbf{System}}
& \multicolumn{4}{c|}{\textbf{Development}}
& \multicolumn{4}{c}{\textbf{Evaluation}} \\
\cmidrule(lr){2-5} \cmidrule(lr){6-9}
& \textbf{L-EER} & \textbf{S-EER} & \textbf{WER} & \textbf{UAR}
& \textbf{L-EER} & \textbf{S-EER} & \textbf{WER} & \textbf{UAR} \\
\midrule
Orig. & 7.34  & --    & 1.80  & 69.08 & 3.41  & --    & 1.84 & 71.06 \\
B2    & 8.15  & 4.23  & 10.44 & 55.66 & 5.62  & 3.72  & 9.96 & 53.49 \\
B3    & 26.03 & 16.32 & 4.31  & 38.08 & 19.06 & 14.24 & 4.31 & 35.24 \\
B4    & 19.73 & 15.28 & 6.16  & 41.97 & 14.64 & 11.13 & 5.90 & 42.78 \\
B5    & 30.94 & 20.05 & 4.91  & 40.08 & 25.65 & 17.05 & 4.44 & 38.25 \\
\midrule
Built-in & 33.15 & 24.64 & 4.18 & 41.85 & 28.19 & 17.78 & 4.58 & 40.38 \\
\midrule
Smooth-21\(^{\dagger}\)
& \cellcolor{gray!15}{35.09} & \cellcolor{gray!15}{25.14} & 4.52 & 40.38
& \cellcolor{gray!15}{29.39} & \cellcolor{gray!15}{21.51} & 4.46 & 39.57 \\
\bottomrule
\end{tabular}
\end{table}
\noindent\textbf{Guidance Scale Grid Search on \emph{Libri\_dev}.}
After verifying the roles of the two scales, we use the same 30-utterance
subset to select the operating point of each system. For each of the three
projection modes (Global, Smooth-21, and Token), we test both target-free AR
construction and random-target-conditioned AR construction. In both cases, the
token-prompt condition \(\boldsymbol{\mu}_{\mathrm{prm}}\) is extracted
from the source utterance. This gives six configurations in total.

For every configuration, both \(\lambda_s\) and \(\lambda_c\) are swept from
\(0\) to \(3\) with a step size of \(0.1\). The source utterances and, when
applicable, their random-target assignments remain fixed throughout the search.
We first retain configurations with WER below \(5\%\), close to the
baselines, and then choose the one with stronger source-speaker dissimilarity.
The \(5\%\) threshold is an application-dependent design choice. The selected
values are reported in Table~\ref{tab:track1-ablation}. After selection, all
scale pairs are fixed and no further tuning is performed on the complete
development or evaluation sets.

Table~\ref{tab:track1-ablation} reports the development-set ablation results, while Table~\ref{tab:track1-semi} compares the selected Smooth-21 system with the official baselines and built-in anonymization.

\noindent\textbf{Effect of Random-Target Conditioning.}
Table~\ref{tab:track1-ablation} isolates the effect of the Stage~1
random-target-conditioned construction of
\(\boldsymbol{\mu}_{\mathrm{AR}}\) introduced in
Section~\ref{sec:condition_construction}.
To provide a controlled comparison, we first keep
\((\lambda_s,\lambda_c)\) exactly the same as in each target-free counterpart
and change only the use of random-target (RT) conditioning.
Under this fixed-scale setting, development Lazy-EER increases from
\(33.90\%\) to \(34.44\%\) for Global, from \(34.20\%\) to \(34.79\%\) for
Smooth-21, and from \(33.48\%\) to \(33.80\%\) for Token. Thus, all three projection modes achieve higher privacy with comparable utility
when RT conditioning is the only changed factor.

When the guidance scales are further tuned for the RT systems, Lazy-EER
increases to \(34.61\%\), \(35.09\%\), and \(34.72\%\) for Global,
Smooth-21, and Token, respectively. These results confirm that the benefit of
random-target conditioning is not caused by different hyperparameter choices:
introducing an unrelated target context itself consistently reduces
source-speaker dependence in the AR-generated part
\(\boldsymbol{\mu}_{\mathrm{AR}}\).
We therefore retain random-target-conditioned AR construction in the selected
system used for the subsequent semi-informed comparison.

\noindent\textbf{Effect of the Projection Method.} Table~\ref{tab:track1-ablation} also evaluates the projected speech-generation
rule introduced in Section~\ref{sec:projected_cfg}. The three configurations without random-target context use the same
target-free construction of
\(\boldsymbol{\mu}_{\mathrm{AR}}\) and the same source-derived
\(\boldsymbol{\mu}_{\mathrm{prm}}\) as the built-in Seed-VC anonymization
mode. Their main difference is therefore in Stage~2. The built-in mode uses the
complete token condition with prompt and style set to zero, whereas the
proposed systems additionally use source prompt and style to estimate the
source-conditioning direction and then remove the aligned part of token
guidance.

All three projection variants improve development Lazy-EER over the built-in
mode: Global, Smooth-21, and Token reach \(33.90\%\), \(34.20\%\), and
\(33.48\%\), respectively, compared with \(33.16\%\) for built-in
anonymization. Their WER remains close to that of the built-in system. This indicates that the privacy gain is not produced only by random-target
conditioning, suppressing the source-aligned component during speech
generation is itself beneficial. Smooth-21 gives the strongest privacy result
among the three projection granularities.

\noindent\textbf{Overall Comparison with Baselines.} Table~\ref{tab:track1-semi} compares the selected
random-target-conditioned Smooth-21 system with the official baselines and the
built-in Seed-VC anonymization mode. On the evaluation set, Smooth-21 reaches a
Lazy-EER of \(29.39\%\) and a Semi-EER of \(21.51\%\), both higher than the
corresponding results of the built-in anonymization mode. It also exceeds the
strongest official baseline B5 in both evaluation EER measures, while
maintaining a comparable WER.
Smooth-21 also achieves the highest development EERs, confirming its
effectiveness against a semi-informed attacker while preserving competitive
utility.

\section{Conclusion and Future Work}

We presented a training-free, two-stage voice anonymization method built on
Seed-VC v2. The ablation results show that random-target-conditioned AR construction
reduces source-speaker dependence in
\(\boldsymbol{\mu}_{\mathrm{AR}}\), while source-conditioning-aware
projected CFG further suppresses source-aligned information carried by the
complete token guidance during speech generation. Together, the two stages improve
speaker privacy while maintaining competitive linguistic and emotion utility
on VPC 2026 Track~1.
The current study is limited to English anonymization. Future work will extend
the method to VPC 2026 Track~2 and investigate its performance in multilingual anonymization.

\bibliographystyle{IEEEtran}
\bibliography{references}

\end{document}